\documentclass[11pt,a4paper]{article}
\pdfoutput=1  
\usepackage{amsmath,amsfonts,amssymb,amsbsy}
\usepackage{graphicx}
\usepackage{color}
\usepackage{booktabs}
\usepackage{alpha}
\usepackage{macros_alpha}
\usebiblio{latticen}

\renewcommand{\strange}{\mathrm{strange}}

\newcommand{\PCAC}{{\mathrm{PCAC}}}

\begin{document}

\preprintno{%
TCDMATH 16-05\\
MS-TP-16-11\\
CP3-Origins-2016-015 DNRF90\\
DIAS-2016-15
}

\title{%
Non-perturbative renormalization of the axial current in $\nf = 3$ lattice QCD
with Wilson fermions and tree-level improved gauge action
}

\collaboration{\includegraphics[width=2.8cm]{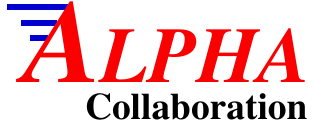}}

\author[trin]{John~Bulava}
\author[ode,esp]{Michele~Della~Morte}
\author[wwu]{Jochen~Heitger}
\author[wwu]{Christian~Wittemeier}

\address[trin]{School~of~Mathematics, Trinity~College, Dublin~2, Ireland}
\address[ode]{CP$^3$-Origins, University of Southern Denmark, Campusvej 55, 5230 Odense M, Denmark}
\address[esp]{IFIC and CSIC, Calle Catedr\'atico Jos\'e Beltran 2, 46980~Paterna, Valencia, Spain}
\address[wwu]{Institut~f\"ur~Theoretische~Physik, Universit\"at~M\"unster, Wilhelm-Klemm-Str.~9, 48149~M\"unster, Germany}

\begin{abstract}
We non-perturbatively determine the renormalization factor of the axial vector
current in lattice QCD with
$\nf=3$ flavors of Wilson-clover fermions and the tree-level Symanzik-improved
gauge action.
The (by now standard) renormalization condition is derived from the massive
axial Ward identity and it is imposed among Schr\"{o}dinger functional states
with large overlap on the lowest lying hadronic state in the pseudoscalar
channel, in order to reduce kinematically enhanced cutoff effects.
We explore a range of
couplings relevant for simulations at lattice spacings
of $\approx 0.09\,\Fm$ and below.
An interpolation formula for $\za(g_0^2)$, smoothly connecting
the non-perturbative values to the 1-loop expression, is provided
together with our final results.
\end{abstract}


\maketitle
\section{Introduction}

It is well known that chiral symmetry is explicitly
broken in the Wilson lattice regularization of QCD~\cite{Wilson}.
As a consequence of that, the isovector axial current does not satisfy the
continuum Ward-Takahashi identities.
These can be restored up to cutoff effects by a finite renormalization
of the axial vector current~\cite{Bochicchio:1985xa}.
Previous computations by the ALPHA Collaboration
in the quenched~\cite{Luscher:1996jn} and in the two-flavor dynamical
cases~\cite{DellaMorte:2005rd,DellaMorte:2008xb} have shown that at the lattice spacings $a$
typically simulated ($0.04\;{\rm fm} \; \lesssim a \lesssim \; 0.1\;{\rm fm}$), this renormalization
factor $\za$ differs significantly from its (1-loop) perturbative estimate.
Since it is required for the computation of pseudoscalar decay constants and
thus, e.g., in the scale setting procedure (through $F_\pi$ or $F_{\rm K}$,
as done in~\cite{lambda:nf2} for $N_{\rm f}=2$ and launched
in~\cite{Bruno:2014jqa} for $N_{\rm f}=2+1$),
as well as in the computation of light~\cite{DellaMorte:2005kg,lambda:nf2}
and heavy~\cite{DellaMorte:2006cb,Bernardoni:2013xba} quark masses, it is
of paramount importance to determine $\za$ non-perturbatively.

Here we report about a non-perturbative determination of $\za$ in
lattice QCD with $\nf=3$ mass-degenerate flavors
of Wilson-clover fermions and the tree-level Symanzik-improved gauge
action~\cite{Luscher:1984xn}.
For a calculation in the three-flavor theory with stout-smeared quarks and
RG-improved Iwasaki gluon action, see Ref.~\cite{Ishikawa:2015fzw}.

The improvement coefficient $\csw$ of lattice QCD with $\nf=3$ O$(a)$
improved Wilson fermions and tree-level Symanzik-improved gauge
action has been non-perturbatively tuned in~\cite{Bulava:2013cta}.
Our computation of $\za$, a preliminary account of which was already given
in~\cite{Bulava:2015vja},
is performed with Schr\"odinger functional boundary conditions,
and we use the same method adopted in~\cite{DellaMorte:2008xb} for the
$\nf=2$ case.
In particular, the normalization
condition exploits the full, massive axial Ward identity in order to reduce
finite quark mass uncertainties in the evaluation of $\za$.
Correlators are built using optimized
boundary wave-functions such that cutoff effects due to excited state
contributions are suppressed.
The setup in the present work concerning
the simulation parameters and the choice of boundary interpolating fields
(wave-functions) is the same as the one recently employed
for the computation of the improvement coefficient
$c_{\rm A}$~\cite{Bulava:2015bxa}.

We discuss the relevant equations for the normalization condition in Section~2
and provide some simulation details in Section~3. Numerical results and the
final interpolation formula are presented in Section~4,
together with a discussion of residual systematic effects.
Section~5 contains our summary.
\section{Renormalization condition}\label{s:renm}

The condition that we choose in order to normalize the axial current
has been originally introduced for the case of two dynamical fermions in~\cite{DellaMorte:2005rd}.
In this section we give a short account of its derivation. More details can be
found in the quoted paper.

The Partially Conserved Axial Current (PCAC) relations are the set of (infinite)  Ward identities derived
performing a chiral rotation of the quark fields. By restricting the transformation
to a region $R$, one can derive different operator relations depending
on the particular choice of composite fields inserted internally and externally w.r.t.\ the
region $R$. If the  axial
current~$A_\nu^b(y)$ is chosen as internal operator,  the
resultant identities can be cast in the integrated form~\cite{Luscher:1996sc}
\begin{equation}
  \label{e:ward1}
  \int_{\partial R} \dd{}{\sigma_\mu(x)}
\mvl{A_\mu^a(x) A_\nu^b(y) \opext} - 2 m \int_R \dd{4}{x} \mvl{P^a(x) A_\nu^b(y) \opext}  = i \epsilon^{abc} \mvl{V_\nu^c(y) \opext},
\end{equation}
where $a,\,b$ and $c$ are flavor indices in a SU(2) sub-group of the chiral group SU(3).
In the equation above, $R$ (containing $y$) is still arbitrary and $\opext$ is
an operator built from fields outside $R$. $V_\nu^c$ is the
isovector vector current.
Further specifying $R$ as the spacetime
volume between two space-like hyperplanes and
setting $\nu = 0$, after contracting the flavor indices~$a$ and
$b$ with $\epsilon^{abc}$, one arrives at
\begin{multline}
  \label{e:ward2}
  \int \dd{3}{\vecx} \dd{3}{\vecy} \epsilon^{abc} \mvl{A_0^a(x) \, A_0^b(y) \, \opext} \\
  {} - 2 m \int \dd{3}{\vecx} \dd{3}{\vecy} \int_{y_0}^{x_0} \dd{}{x_0} \epsilon^{abc} \mvl{P^a(x) \, A_0^b(y) \, \opext}
  \\ = i \int \dd{3}{\vecy} \mvl{V_0^c(y) \, \opext}
\end{multline}
with $x_0 > y_0$ defining the hyperplanes. It is clear that the above relation, once considered at the
renormalized level, relates the normalization of the axial current to that of the vector current.
A condition for the latter will be implicitly given below.

We evaluate the identity in \eq{e:ward2} on the lattice with Schr\"odinger functional
boundary conditions (periodic in space, Dirichlet in
time)~\cite{SF:LNWW,SF:stefan1} with vanishing background field. The
source operator $\opext$ is expressed in terms of the quark fields
$\zeta$ and $\zetaprime$ at the boundaries $x_0 = 0$ and $x_0 = T$ as
\begin{equation}
  \label{e:opext}
  \opext = -\frac{1}{6 L^6} \epsilon^{cde} \opprime{}^d {\mathcal O}^e
\end{equation}
with
\begin{equation}
  \label{e:sourceops}
  {\mathcal O}^e             = a^6 \sum_{\vecu, \vecv} \zetabar(\vecu) \, \gamma_5 \, \frac{\tau^e}{2} \, \omega(\vecu-\vecv) \, \zeta(\vecv)  \;\; \text{and}\;\;
  \opprime{}^d      = a^6 \sum_{\vecu, \vecv} \zetabarprime(\vecu) \, \gamma_5 \, \frac{\tau^d}{2} \, \omega(\vecu-\vecv) \, \zetaprime(\vecv)\;.
\end{equation}
The wave-function~$\omega$ is optimized in order to excite states with a large
projection on the pseudoscalar ground state. Its construction is detailed in
\cite{Bulava:2015bxa}. The free index~$c$ in \eq{e:opext} is
contracted with the free index in \eq{e:ward2}. In this case, the
term on the right-hand side involving the isospin charge density can be
simplified to the boundary-to-boundary correlator
\begin{equation}
  F_1  = -\frac{1}{3 L^6} \mvl{\opprime{}^a {\mathcal O}^a}
\end{equation}
up to $\Oasq$, as it has been shown in~\cite{Luscher:1996jn,DellaMorte:2005rd} by using
isospin symmetry.

After replacing all terms by their improved and renormalized lattice counterparts, the Ward identity can be written as
\begin{equation}
  \label{e:ward3}
  \za^2 \left( 1 + \ba \, a \mq \right)^2 \left[ \faa^{\rm I}(x_0, y_0) - 2 m \cdot \fpatilde^{\rm I}(x_0, y_0) \right]  = F_1,
\end{equation}
with the improved correlation functions
\begin{eqnarray}
  \label{e:faai}
  \faa^{\rm I}(x_0, y_0) && = \faa(x_0, y_0) + a \ca \left[ \tilde\partial_{x_0} \fpa(x_0, y_0) + \tilde\partial_{y_0} \fap(x_0, y_0) \right] \nonumber \\
                    && \qquad +\,a^2 \ca^2 \, \tilde\partial_{x_0} \tilde\partial_{y_0} \fpp(x_0, y_0), \\
  {\rm and} \quad
  \label{e:fpaitilde}
  \fpatilde^{\rm I}(x_0, y_0) && = a \sum_{x'_0=y_0}^{x_0} w(x'_0) \left[ \fpa(x_0, y_0) + a \ca \, \partial_{y_0} \fpp(x_0, y_0) \right],
\end{eqnarray}
where $\tilde\partial$ denotes the central difference operator and
$F_{XY}(x_0, y_0)$ with $X, Y \in \{A_0, P\}$ reads
\begin{equation}
  \label{e:fxy}
  F_{XY}(x_0, y_0)  = -\frac{a^6}{6 L^6} \sum_{\vecx, \vecy} \epsilon^{abc} \epsilon^{cde} \mvl{\opprime{}^d X^a(x) Y^b(y) {\mathcal O}^e}\;,
\end{equation}
and
\begin{equation}
  w(x'_0)  = \begin{cases}
                1/2 & \text{if $x'_0 = y_0$ or $x'_0 = x_0$} \\
                1   & \text{if $y_0 < x'_0 < x_0$}
              \end{cases}
\end{equation}
implements the trapezoidal rule.
%
%
In \eq{e:ward3} the bare quark mass $m$ is defined through the PCAC relation, while $m_{\rm q}$
is the bare subtracted quark mass. The mass-dependent improvement term
proportional to $\ba$ will be neglected from here on, since we will
impose the renormalization condition at vanishing mass.
Any mistuning will result in $\Or(am)$ effects, as
effects of $\Or(\Lambda_{\rm QCD}m)$ are explicitly removed by using
the massive Ward identity.
Our final renormalization
condition thus reads
\begin{equation}
  \label{e:renormalizationcondition}
  \za  = \lim_{m \to 0} \left[ \frac{F_1}{\faa^{\rm I}(x_0, y_0) - 2 m \cdot \fpatilde^{\rm I}(x_0, y_0)} \right]^{\frac{1}{2}}.
\end{equation}
In order to maximize the distance between the insertion points and to keep
this distance physical (once $L$ is fixed),
we choose $x_0 = \frac{2}{3} T$ and $y_0 = \frac{1}{3} T$.

\begin{figure}
  \centering
  \includegraphics{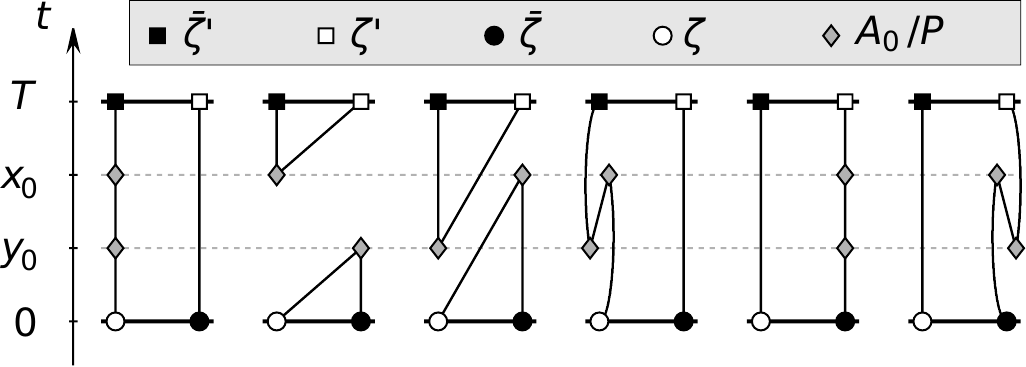}
  \caption{The six non-vanishing Wick contractions contributing to the correlation
    functions~$F_{XY}$ with sources ${\mathcal O}^\mathrm{ext}$ on the boundaries and two
    insertions of the operators $X$ and $Y$ in the bulk, see \protect\eq{e:fxy}, taken
    from~\cite{DellaMorte:2005rd}.}
  \label{f:contractions}
\end{figure}

Except for $F_1$,  only
correlators of the form given in \eq{e:fxy} appear in \eq{e:renormalizationcondition}.
After working out the Wick contractions, one finds that only six
diagrams contribute to those. They are depicted in
\fig{f:contractions}. Two of them are disconnected, and as showed in Appendix~A of
Ref.~\cite{DellaMorte:2005rd}, they only give rise to $\Oasq$ contributions and
vanish in the massless continuum limit. By omitting them and taking only the
connected contractions, one obtains an alternative renormalization
condition. The corresponding renormalization factor is denoted by
$\zacon$ in the following. Although our preferred definition remains the
one including all quark-contractions, $\zacon$ offers the possibility
to further check the smooth
dependence of $\za$ on the gauge coupling, as a consequence
of the constant physics setup, as well as the smooth  $\Oasq$ convergence
of different renormalization conditions from the explored non-perturbative
region to the perturbative regime.

An alternative renormalization condition, using 
the Schr\"odinger functional with chirally rotated boundary conditions,
has been  recently proposed and tested
in perturbation theory~\cite{Brida:2016rmy}
and non-perturbatively in the quenched and in the $\Nf=2$ cases~\cite{Sint:2010xy,Brida:2014zwa}.
The main advantage of the approach being that it entails automatic O($a$)
improvement, it seems very promising and, in two-flavor QCD, turned out
to yield more precise results than with standard Schr\"odinger functional
boundary conditions~\cite{Brida:2014zwa}.
\section{Simulation details}\label{s:sims}
The ensembles used in this study coincide with those considered in~\cite{Bulava:2015bxa},
and algorithmic details can be found there.
In a few cases the number of configurations has actually been enlarged, and in addition
we generated a new ensemble (at $L/a=14$) with the purpose of better constraining the final
parameterization of the renormalization constant in the region
where the dependence on the gauge coupling is strongest.

Our three-flavor lattice QCD simulations with Schr\"{o}dinger functional
boundary conditions used the \texttt{openQCD}
code\footnote{\url{http://luscher.web.cern.ch/luscher/openQCD/}}%
of Ref.~\cite{algo:openQCD}.
In order
to ensure a smooth dependence of the renormalization constant $Z_{\rm A}$ on the
bare gauge coupling, we have approximately fixed a constant physics condition
by setting  $L \approx 1.2\,\mathrm{fm}$.
That is achieved
by beginning with a particular pair of $g_0^2$ and
$L/a$ ($\beta =6/g_0^2 = 3.3$ at $L/a = 12$ here) and then choosing
the bare couplings for
subsequent smaller lattice spacings according to the universal 2-loop
$\beta$-function.
In this way we cover lattice spacings in the range from
$a\approx 0.09\,\Fm$ to $a\approx 0.045\,\Fm$.
At each bare coupling we have tuned the
bare quark mass so that the PCAC mass is  close to zero.
We could check at several lattice spacings that
our determination of $\za$ is insensitive to variations of the (small) quark
mass.
Information about our ensembles, consisting in most cases of several replica per parameter
set, are summarized in
Tab.~\ref{t:sim}. For practical reasons discussed in the
\texttt{openQCD} documentation, our lattices have temporal extents
$T = 3L/2 - a$.\footnote{%
For this work we employ \texttt{openQCD} version 1.2.
This issue has been corrected in the latest version (1.4).}
Since we use an $\Or(a)$ improved setup, this offset is expected to influence  the
determination of $\za$ at $\Or(a^3)$ only.
\begin{table}[h]

\centering

\renewcommand{\arraystretch}{1.25}

\setlength{\tabcolsep}{3pt}

\begin{tabular}{ccclccrrcc}

\toprule

$L^3\times T / a^4$ &&& $\beta$ && $\kappa$ & \#\,REP & \#\,MDU && ID   \\

\midrule

$12^3\times 17$     &&& 3.3     && 0.13652  & 10      & 10240   && A1k1 \\

		    &&&         && 0.13660  & 10      & 12620   && A1k2 \\

\hline

$14^3\times 21$     &&& 3.414   && 0.13690  & 32      & 10360   && E1k1 \\

		    &&&         && 0.13695  & 48      & 13984   && E1k2 \\

\hline

$16^3\times 23$     &&& 3.512   && 0.13700  & 2       & 20480   && B1k1 \\

                    &&&         && 0.13703  & 1       & 8192    && B1k2 \\

                    &&&         && 0.13710  & 3       & 24560   && B1k3 \\

\hline

$16^3\times 23$     &&& 3.47    && 0.13700  & 3       & 29584   && B2k1 \\

\hline

$20^3\times 29$     &&& 3.676   && 0.13700  & 4       & 15232   && C1k2 \\

                    &&&         && 0.13719  & 4       & 15472   && C1k3 \\

\hline

$24^3\times 35$     &&& 3.810   && 0.13712  & 5       & 10240   && D1k1 \\

\bottomrule

\end{tabular}

\caption{
Summary of simulation parameters, number of replica and total number of
molecular dynamics units of our gauge configuration ensembles labeled by
`ID', used for the determination of the renormalization factor $\za$.
Compared to the data basis underlying our earlier computation of $\ca$
in~\cite{Bulava:2015bxa}, the analysis presented here includes the additional $L/a=14$
lattice ensembles \{E1k1, E1k2\}.
Also note that the statistics of ensemble B2k1 have been increased by more
than a factor of three.
}\label{t:sim}

\end{table}

In the context of our earlier computation of $c_{\rm A}$~\cite{Bulava:2015bxa}, we already
estimated the deviation from the, perturbatively implemented,
constant $L$ condition, which is also imposed here, by measuring the
scale-dependent renormalized coupling $\bar{g}_{\rm GF}^2$,
defined in Ref.~\cite{flow:FR}.
The results for this coupling
apply here, too, and can be found in Tab.~2 of Ref.~\cite{Bulava:2015bxa}.
Finally, we also test the dependence of $\za$ on $L$ in physical units directly by
simulating an additional bare coupling at $L/a=16$.
\section{Results}\label{s:results}

We measure
the correlation functions defined in Section~\ref{s:renm}
on each fourth trajectory of length $\tau=2\,\mathrm{MDU}$ so that the spacing
between the measurements is $8\,\mathrm{MDU}$. Only on the A1k2 ensemble,
we use a measurement separation of $2\tau=4\,\mathrm{MDU}$.
The total statistics for the different
ensembles are given in Tab.~\ref{t:sim}.

For diagnostic purposes, we also compute `smoothed'
gauge field observables obtained from the Wilson (gradient)
flow~\cite{flow:ML}. We fix the flow time $t$ by setting $\sqrt{8t}/L = c$ with
$c=0.35$.
The smoothed gauge fields provide a renormalized
definition of the topological charge $Q$, which we use to monitor the topology
freezing; in addition, even at lattice spacings
where topology freezing does not occur, the smoothed topological charge and
action typically possess the largest observed autocorrelation times.
Finally,  the coupling
$\bar{g}_{\rm GF}^2$ of Ref.~\cite{flow:FR}, defined through the Wilson flow,
may be used to monitor the deviation from the constant physics condition,
as it depends only on
the physical lattice size up to cutoff effects. For results on this coupling
we refer again to Tab.~2 in~\cite{Bulava:2015bxa}, where one can see that
$\bar{g}_{\rm GF,0}^2(L)$ (i.e., the gradient flow coupling within the zero
topology sector, see below) varies between 14 and 18 on our ensembles.
That roughly corresponds to a 20\% variation in $L$.
\begin{table}
\centering
\renewcommand{\arraystretch}{1.25}
\setlength{\tabcolsep}{3pt}
\begin{tabular}{ccr@{.}lr@{.}lr@{.}lr@{.}lr@{.}lr@{.}l}
\toprule
ID && \multicolumn{2}{c}{$am_\PCAC$} & \multicolumn{2}{c}{$am_{\PCAC,0}$} & \multicolumn{2}{c}{$\za$} & \multicolumn{2}{c}{$Z_{\mathrm{A},0}$} & \multicolumn{2}{c}{$Z_{\mathrm{A}}^{\mathrm{con}}$} & \multicolumn{2}{c}{$Z_{\mathrm{A},0}^{\mathrm{con}}$} \\
\midrule
A1k1       & & $-0$ & $00234(85)$ & $-0$ & $00367(92)$  & $0$ & $641(13)$  & $0$       & $636(16)$        & $0$ & $809(11)$  & $0$ & $811(12)$  \\
A1k2       & & $-0$ & $01085(71)$ & $-0$ & $01206(81)$  & $0$ & $6432(98)$ & $0$       & $638(11)$        & $0$ & $833(11)$  & $0$ & $835(16)$  \\
           & & $0$  & $0$         & $0$  & $0$          & $0$ & $6424(78)$ & ${\it 0}$ & ${\it 6374(91)}$ & $0$ & $8210(78)$ & $0$ & $8196(96)$ \\
\midrule
E1k1       & & $0$  & $00275(38)$ & $0$  & $00262(49)$  & $0$ & $7148(99)$ & $0$       & $727(14)$        & $0$ & $7705(91)$ & $0$ & $7743(98)$ \\
E1k2       & & $0$  & $00004(33)$ & $-0$ & $00072(41)$  & $0$ & $7151(93)$ & $0$       & $702(13)$        & $0$ & $7892(98)$ & $0$ & $773(18)$  \\
           & & $0$  & $0$         & $0$  & $0$          & $0$ & $7150(68)$ & ${\it 0}$ & ${\it 7136(95)}$ & $0$ & $7792(67)$ & $0$ & $7740(86)$ \\
\midrule
B1k1       & & $0$  & $00565(16)$ & $0$  & $00554(23)$  & $0$ & $7666(47)$ & $0$       & $7602(72)$       & $0$ & $7743(24)$ & $0$ & $7744(43)$ \\
B1k2       & & $0$  & $00494(25)$ & $0$  & $00423(36)$  & $0$ & $7676(71)$ & $0$       & $766(10)$        & $0$ & $7759(44)$ & $0$ & $7779(63)$ \\
B1k3       & & $0$  & $00160(18)$ & $0$  & $00109(21)$  & $0$ & $7521(39)$ & $0$       & $7515(52)$       & $0$ & $7806(30)$ & $0$ & $7757(41)$ \\
           & & $0$  & $0$         & $0$  & $0$          & $0$ & $7595(28)$ & ${\it 0}$ & ${\it 7562(39)}$ & $0$ & $7766(17)$ & $0$ & $7756(27)$ \\
\midrule
B2k1       & & $0$  & $00349(17)$ & $0$  & $00306(23)$  & $0$ & $7456(44)$ & $0$       & $7434(59)$       & $0$ & $7778(27)$ & $0$ & $7793(35)$ \\
\midrule
C1k2       & & $0$  & $00618(14)$ & $0$  & $00610(25)$  & $0$ & $7875(92)$ & $0$       & $789(17)$        & $0$ & $7904(22)$ & $0$ & $7840(40)$ \\
C1k3       & & $-0$ & $00082(12)$ & $-0$ & $00099(13)$  & $0$ & $7771(29)$ & $0$       & $7779(31)$       & $0$ & $7833(27)$ & $0$ & $7841(29)$ \\
           & & $0$  & $0$         & $0$  & $0$          & $0$ & $7780(28)$ & ${\it 0}$ & ${\it 7783(30)}$ & $0$ & $7876(17)$ & $0$ & $7841(23)$ \\
\midrule
{\it D1k1} & & n    & q.          & $-0$ & $002909(72)$ & n   & q.         & ${\it 0}$ & ${\it 7897(19)}$ & n   & q.         & $0$ & $8009(35)$ \\
\bottomrule
\end{tabular}
\caption{%
Summary of results for $\za$.
The (unrenormalized) PCAC quark mass $am_\PCAC$ is computed from the
correlation functions projected to the approximate ground state,
using the non-perturbative result for $\ca(g_0^2)$ from \cite{Bulava:2015bxa}
and averaging the local mass over the central four timeslices.
Here, quantities with the explicit subscript label `0' refer to
results from the analysis restricted to the sector of vanishing
topological charge, whereas in the text we loosely suppress the `0'.
Numbers for ensemble D1k1 ($L/a=24$) are not quoted (`n.q.') for the
case of including all charge sectors in the partition sum, because
owing to an insufficient sampling of all the sectors by our
simulations a reliable error estimation is not possible.
For those ensembles, where simulations at several quark masses were
performed, the renormalization constant has been extrapolated to the
chiral limit ($am_\PCAC,am_{\PCAC,0}\to 0$) taking a weighted average
(i.e., fitting to a constant).
Results in italics enter into the final interpolation formula for
$\za(g_0^2)$, eq.~(\ref{eq:final0}).
}\label{t:res}
\end{table}

As discussed in~\cite{Bulava:2015bxa}, for all simulations and all observables, we find that integrated autocorrelation times
are bounded by $\tau_{\mathrm{max}} \lesssim 200-250\,\mathrm{MDU}$, except for our $L/a=24$
simulations where the charge is frozen.
Since we are practically unable to sufficiently sample all topological sectors at
this finest lattice spacing, we everywhere define observables
restricted to the trivial, i.e., $Q=0$ sector.
Note that the same strategy was followed in our
non-perturbative determination of $c_{\rm A}$ in~\cite{Bulava:2015bxa}.
As shown in Tab.~\ref{t:res}, the projection to the $Q=0$ sector
does not induce a noticeable difference in the final numbers for $\za$,
which is expected since the Ward identities, being operator relations,
 are valid in each topological sector.

Statistical errors are estimated by
applying a
full autocorrelation analysis according to Ref.~\cite{Wolff:2003sm}.

For the wave-functions $\omega$ in eq.~(\ref{e:sourceops}), we
use the same approach adopted in~\cite{Bulava:2015bxa} and
solve for the two largest
eigenvectors of the matrix $F_1(\omega_i', \omega_j)$.
These normalized eigenvectors have a
well-defined continuum limit along our line of constant physics in
parameter space, as long as the wave-functions depend on physical scales only.
Since we do not observe any significant lattice spacing dependence for
them, we fix these eigenvectors to the values calculated on the
B1k2 ensemble ($L/a=16$, $\beta=3.512$, $\kappa=0.13703$) and regard that
as part of our choice of the renormalization condition.
The effective masses of the correlation function $f_{\rm P}$, after taking the
inner product with the eigenvectors in wave-function space,
indicate clearly distinct signals, providing evidence that those
effectively maximize the overlap with the ground and first excited states
(see Fig.~2 in~\cite{Bulava:2015bxa}).
Notice that, as opposed to the case of the computation of the improvement
coefficient $c_{\rm A}$, here we only need the wave-function projecting onto
the ground state.\footnote{%
Explicit expressions for the basis wave-functions entering in this analysis,
as well as for the resultant eigenvector projecting onto the (approximate)
ground state can be found in Ref.~\cite{Bulava:2015bxa}.
}

In the following, we restrict the discussion of systematic effects to
observables projected to the sector of vanishing topological charge:
$am_\PCAC\equiv am_{\PCAC,0}$, $Z_{\rm A}\equiv Z_{\rm A,0}$.
As mentioned above, those are the ones entering our final results.
In any case, the `un-projected' quantities, where they can be properly
estimated, display the same qualitative features.

\begin{figure}[t!]
 \centering
  \includegraphics[width=0.75\textwidth]{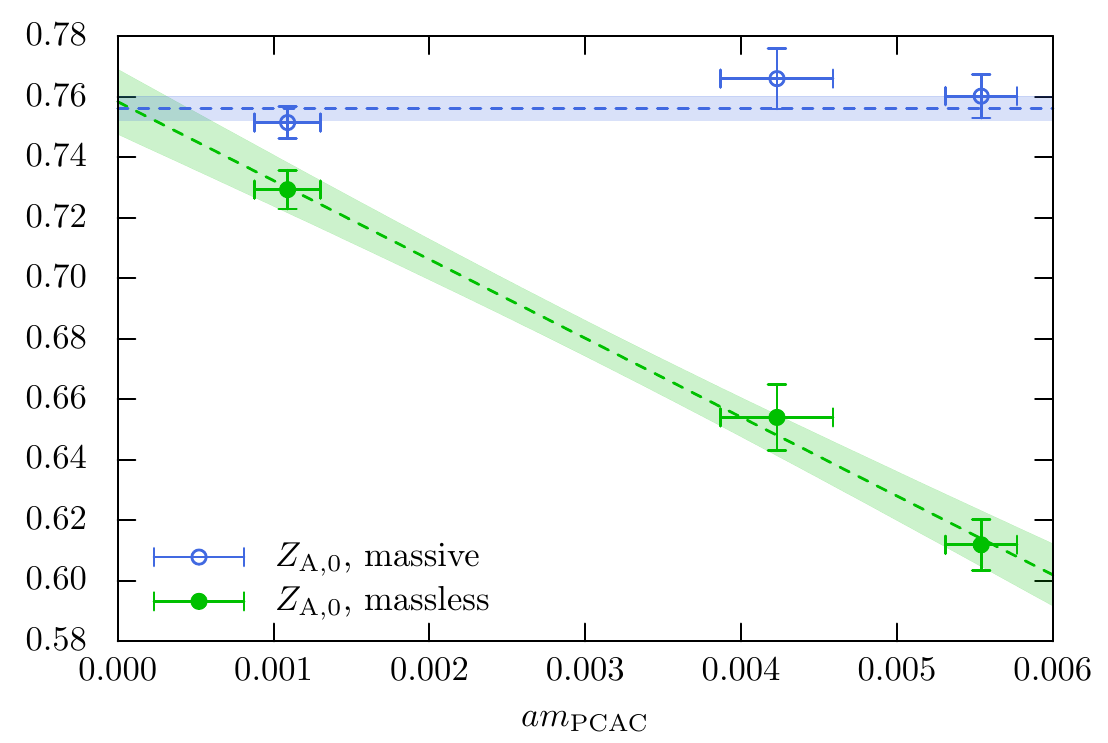}
  \caption{Chiral extrapolations for $Z_{\rm A}=Z_{\rm A,0}$ imposing the
  massive and massless conditions on ensembles
  \{B1k1, B1k2, B1k3\} at $L/a=16, \, \beta=3.512$.
  The data in the massive case, for which a slope is found to be
  statistically insignificant, are fitted to a constant.}
  \label{f:massdep}
\end{figure}

As expected from the discussion in Section~\ref{s:renm},
for the massive normalization condition the data exhibit very little
dependence on the quark mass, which implies that uncertainties due to the
position of the critical mass do not affect the determination of $\za$.
We illustrate the chiral extrapolation $am_{\rm PCAC}\to 0$ at $\beta=3.512$
in Fig.~\ref{f:massdep}.
For this normalization condition, the slope in $am_{\rm PCAC}$
is consistent with zero,
whereas the estimate of $\za$  from the massless Ward identity
(i.e., by setting directly $m=0$ in eq.~(\ref{e:renormalizationcondition}))
changes by 25\% in the very small mass range displayed.
At all the other gauge couplings the situation is very similar, with
$\za$ from the massive Ward identity definition being mass-independent
within errors, as evinced by Tab.~\ref{t:res}.
In fact, performing fits to a constant over the considered mass range
yields suitably small
$\chi^2$- and very reasonable goodness-of-fit-values and, therefore, is
fully consistent with our data.
Moreover, the slopes, which would come out of linear fits, are compatible
with zero within one standard deviation for most ensembles
(and within $1.5\sigma$ for all), and their magnitude is such that we do not
expect them to have any relevant impact on our final results and errors.
All this serves as a further justification of this chiral extrapolation
procedure.

Near the continuum, and in the $\Or(a)$ improved theory, the dependence of
renormalization factors on the lattice extent is
expected to be an $\Or\left((a/L)^2\right)$ effect, hence deviations from the line of constant physics
should affect our determination of $\za$ by the same amount.
The B2k1 ensemble has been generated exactly with the purpose of checking these effects,
as it differs from the ensembles in the B1 series by a 6\% change in $L$.
The value of  $\za$ determined there
lies within one standard deviation from the chirally extrapolated value for the B1 series,
and we are therefore confident
that (small) variations of $L$ do not produce significant shifts in our estimates of $\za$.

Our final results for $\za$, after chiral extrapolations via performing fits
to a constant at each $\beta$ as explained before,
are shown in Fig.~\ref{f:final} as a function of $g_0^2$.
One can see that the data lie on a smooth curve, which we describe by
performing a Pad\'{e} fit, producing the expression:
\begin{figure}[t!]
 \centering
  \includegraphics[width=0.75\textwidth]{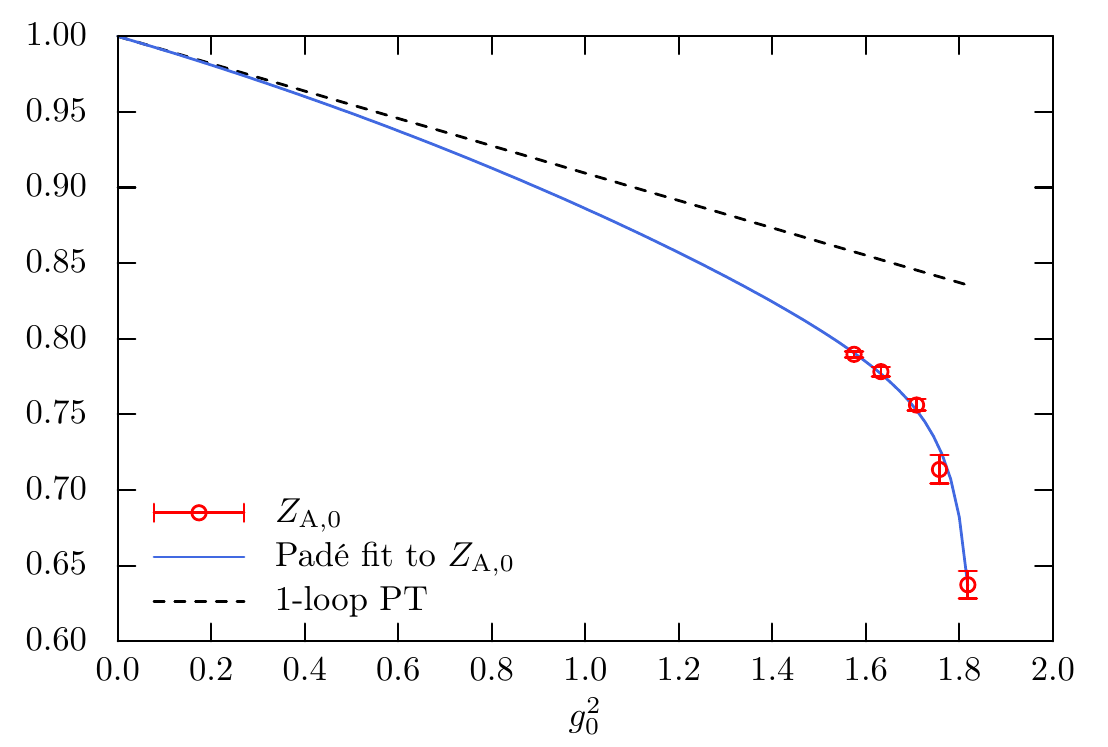}
  \caption{Final results and interpolation for $\za(g_0^2)=Z_{\rm A,0}(g_0^2)$.}
  \label{f:final}
\end{figure}
\begin{equation}
\za(g_0^2) = 1 - 0.090488 \, g_0^2 \times
\frac{ 1 - 0.29026 \, g_0^2  -0.12881 \, g_0^4}{1 - 0.53843 \, g_0^2 }\;.
\label{eq:final0}
\end{equation}
The associated $\chi^2/{\rm d.o.f.}$ is 1.71.
Notice that the 1-loop perturbative formula
$\za(g_0^2) = 1 - 0.090488\,g_0^2$, extracted for our gauge action from the
results of the calculation in~\cite{Aoki:1998ar}, is imposed
as asymptotic constraint. Errors at the directly simulated $\beta$-values
decrease in relative size from about 1.4\% at the coarsest lattice spacing
(corresponding to $\beta=3.3$) to about 2.4\textperthousand~at $\beta=3.81$.
It is interesting to observe that for $\beta \geq 5.5$ our interpolation
formula results are consistent with those obtained in~\cite{Constantinou:2014fka} for $Z_{\rm A}$  with 
3 flavors of stout link non-perturbative clover fermions using a different scheme (RI'-MOM).

\begin{figure}[t!]
 \centering
  \includegraphics[width=0.75\textwidth]{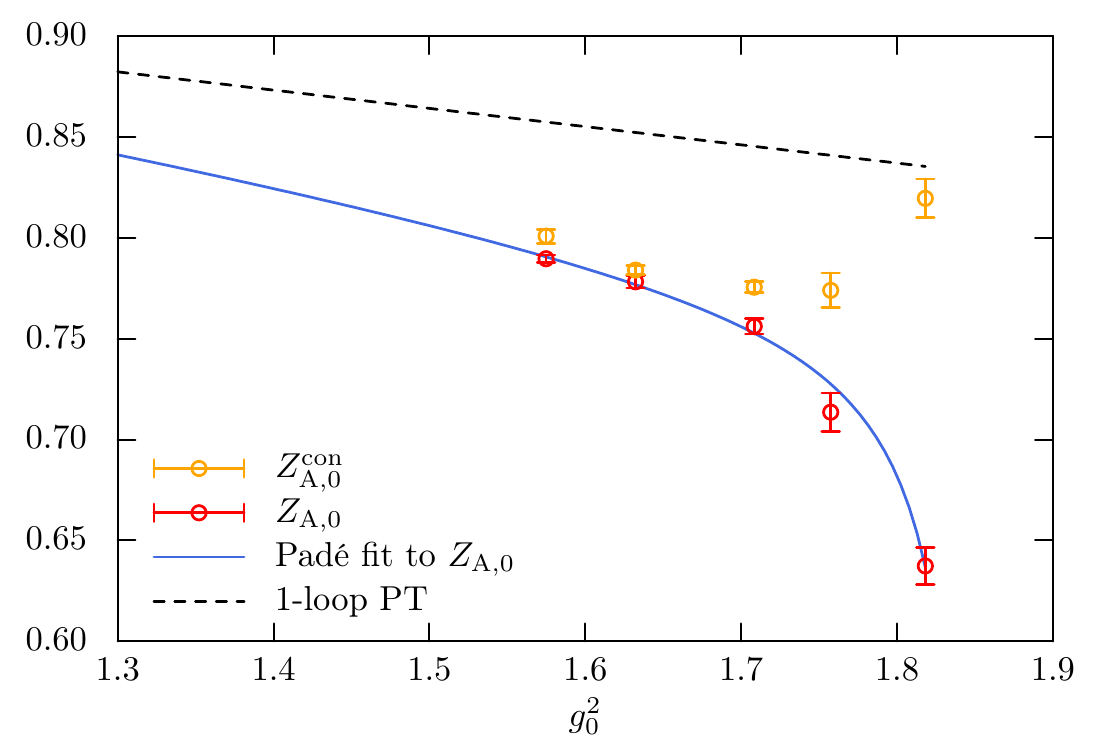}
  \caption{Comparison of $\za(g_0^2)$ and $\za^{\rm con}(g_0^2)$ for
  $1.3 \leq g_0^2 \leq 1.82$.}
\label{f:convsfull}
\end{figure}

\begin{figure}[t!]
 \centering
  \includegraphics[width=0.75\textwidth]{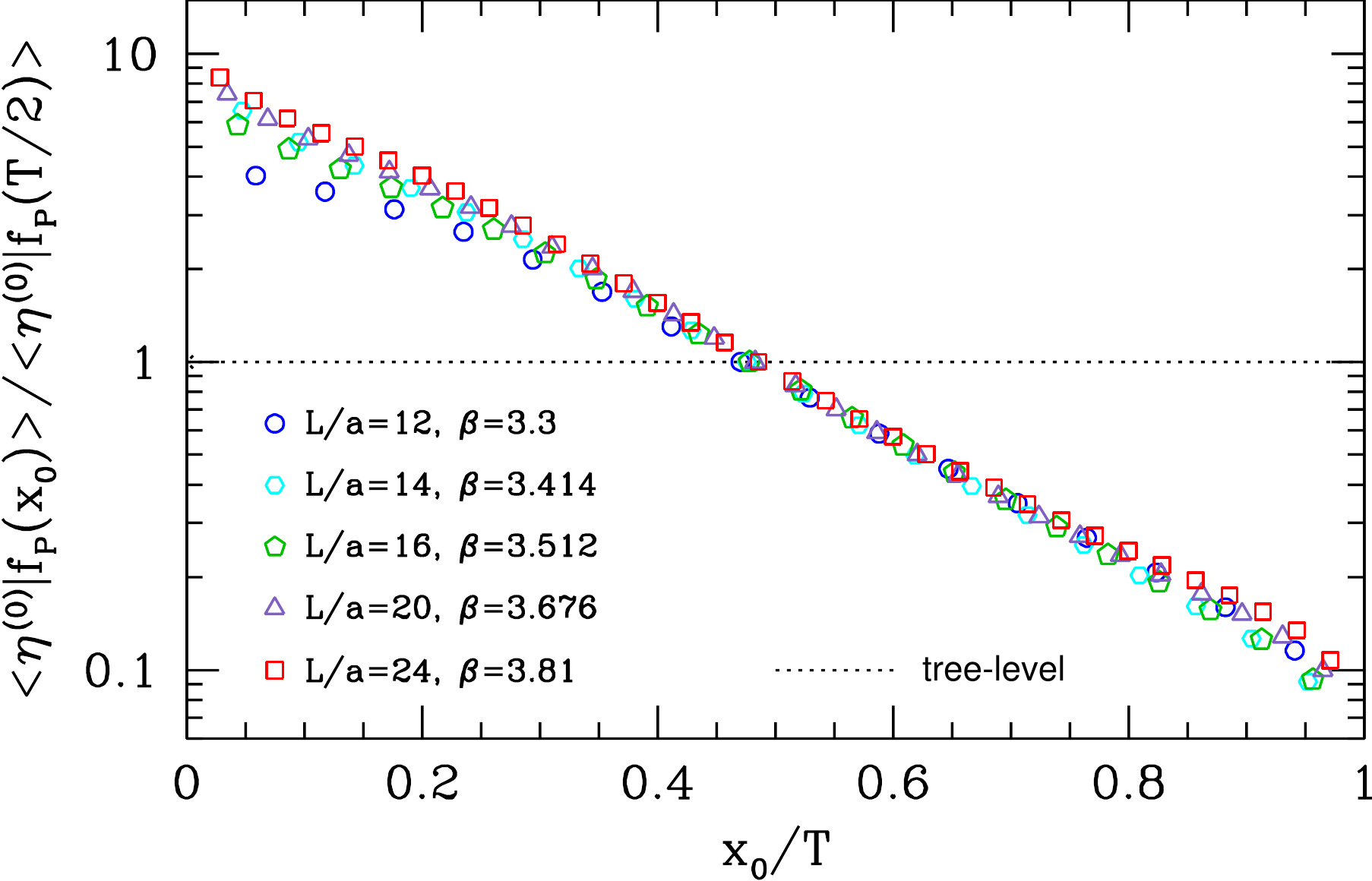}
  \caption{The correlator $f_{\rm P}(x_0)$ projected to the approximate
  ground state (i.e., by taking the inner product with the associated
  eigenvector denoted by $\eta^{(0)}$ in~\cite{Bulava:2015bxa}), for
  the different values of the gauge coupling along the chosen line of
  constant physics.
  Errors are of the size of the symbols and hence suppressed
  for better readability.}
\label{f:Palla}
\end{figure}

In our setup, an alternative definition of $\za$ can be obtained by dropping the disconnected diagrams
since, as discussed in Section~\ref{s:renm}, those are expected to contribute at $\Or(a^2)$ only.
The results for the corresponding  $\za^{\rm con}$ are reproduced in
Fig.~\ref{f:convsfull}, where they
are also put in comparison to the interpolation formula in
eq.~(\ref{eq:final0}) for $1.3 \leq g_0^2 \leq 1.9$.
The difference between the two definitions amounts to a cutoff effect,
and we could indeed explicitly check that it vanishes even faster than $a^2$.
Compared to the $N_{\rm f}=2$ case in~\cite{DellaMorte:2005rd},
we observe a much smoother,
with the exception\footnote{%
Let us remark in this context that at the coarsest lattice spacing of
$a\approx 0.09\,\Fm$ ($\beta=3.3$) substantial cutoff effects
(e.g., for Wilson flow observables) were also encountered in large-volume
$(2+1)$-flavor QCD simulations~\cite{Bruno:2014jqa}, with the same setup of
non-perturbatively improved Wilson fermions in the sea and the
L\"uscher-Weisz action for the gluons as used here.
}%
of the point at $a\approx 0.09\,\Fm$
almost flat, dependence of $\za^{\rm con}$ on $g_0^2$ at the lattice spacings
considered.
We ascribe that to the choice of the kinematical setup ($L \approx$ 1.2 fm) and to the
approximate isolation of the ground state in the correlation functions involved, which we adopted
following the suggestion put forward in~\cite{DellaMorte:2008xb} in order to minimize intermediate-distance cutoff effects.
Indeed, as depicted in Fig.~\ref{f:Palla}, we see a rather slow decay in time of the correlation
function $f_{\rm P}(x_0)$ approximately projected to the ground state, and very moderate
lattice artifacts. The slope corresponds to an effective mass which is
smaller than 0.3 in units of the lattice spacing, and the
approximate linear behavior of the correlators in the plot implies
that they are
dominated by a few states (most likely one, except at very short
distances), all with energies well below the cutoff scale.
This is in contrast with the quite strong time-dependence observed 
in~\cite{Aoki:2010wm} for the correlators entering the definition of $\za$.
In that case, corresponding to 3 flavors of non-perturbatively
improved Wilson fermions with Iwasaki gauge action,
rather small volumes and wall-sources (without attempting 
to project on the ground state) have in fact been used.
\section{Conclusions}\label{s:concl}
In this work we have non-perturbatively determined $\za(g_0^2)$, the
renormalization factor of the axial vector current
matrix elements, in lattice QCD with $\nf=3$ flavors of Wilson quarks,
non-perturbative $\csw$~\cite{Bulava:2013cta} and the tree-level
Symanzik-improved gauge action.
The renormalization condition is chosen such that the Ward identities
are restored up to O$(a^2)$ at finite lattice spacing.
The main result is the parameterization of $\za(g_0^2)$,
eq.~(\ref{eq:final0}), valid for bare couplings below $g_{0}^2 \approx 1.8$
(or, equivalently, for lattice spacings $a \lesssim 0.09\,\Fm$).

As the range of lattice spacings covered in this work matches those of the
large-volume $\nf=2+1$ flavor QCD ensembles of gauge field configurations
currently being generated in dynamical simulations with the same lattice
action~\cite{Bruno:2014jqa}, the present calculation (together with the
determination of the improvement coefficient $\ca$ in~\cite{Bulava:2015bxa})
is a useful ingredient in the computation of quark masses as well as of
pseudoscalar meson decay constants, which can be used to convert lattice
spacings to physical units and are of great phenomenological interest by
their own.

\vskip 0.2cm

\noindent
{\bf Acknowledgments.}
We thank Rainer Sommer for helpful discussions.
This work is supported by the
grant {HE 4517/3-1} (J.~H.\ and C.~W.) of the Deutsche Forschungsgemeinschaft.
We gratefully acknowledge the computing time granted by the John von Neumann
Institute for Computing (NIC) and provided on the supercomputer JUROPA at
J\"{u}lich Supercomputing Centre (JSC).
Computer resources were also provided by DESY, Zeuthen (PAX cluster),
the CERN `thqcd2' QCD HPC installation, and the ZIV of the University of
M\"{u}nster (PALMA HPC cluster).



\end{document}